\documentstyle[12pt,epsfig]{article}

\def\Title#1{\begin{center} {\Large {\bf #1} } \end{center}}

\begin{document}

\Title{Cosmic Microwave Background: Past, Future, and Present}

\bigskip\bigskip


\begin{raggedright}  

{\it Scott Dodelson\index{Dodelson, Scott}\\
NASA/Fermilab Astrophysics Center \\
P.O. Box 500\\
Batavia, IL 60510}
\bigskip\bigskip
\end{raggedright}

\section{Introduction}

Anisotropies in the Cosmic Microwave Background (CMB)
carry an enormous amount of information about the early
universe. The anisotropy spectrum depends sensitively on
close to a dozen cosmological parameters, some of which 
have never been measured before. Experiments over the next
decade will help us extract these parameters, teaching us
not only about the early universe, but also about physics
at unprecedented energies. We are truly living in the Golden
Age of Cosmology. 

One of the dangers of the age is that we are tempted
to ignore the present data and rely too much 
on the future. This would be a shame, for hundreds of
individuals have put in countless [wo]man-years building
state-of-the-art instruments, making painstaking observations
at remote places on and off the globe. It seems unfair to ignore all
the data that has been taken to date simply because there will
be more and better data in the future. 

In this spirit, I would like to make the following claims:

\begin{itemize}

\item We understand the theory of CMB anisotropies.

\item Using this understanding, we will be able to extract
from {\it future} observations extremely accurate measurments of
about ten cosmological parameters.

\item Taken at face value, {\it present} data determines one of these
parameters, the curvature of the universe.

\item The present data is good enough that we should believe these
measurements.

\end{itemize}

The first three of these claims are well-known and difficult to argue with;
the last claim is more controversial, but I will present evidence for it
and hope to convince you that it is true. If you come away a believer, then
you will have swallowed a mouthful, for the present data strongly suggest
that the universe has zero curvature. If you believe this data, then you believe
that (a) a fundamental prediction of inflation has been verified and (b) 
since astronomers do not see enough matter to make the universe flat, roughly
two-thirds of the energy density in the universe is of some unknown form.

\section{Anisotropies: The Past}

When the universe was much younger, it was denser and hotter. When the
temperature of the cosmic plasma was larger than about $1/3$ eV, there were
very few neutral hydrogen atoms. Any time a free electron and proton came together
to form hydrogen, a high energy ($E>13.6$ eV) photon was always close enough to
immediately dissociate the neutral atom. After the temperature dropped beneath a
$1/3$ eV, there were no longer enough ionizing photons around, so virtually
all electrons and protons combined into neutral hydrogen. This transition --
called recombination --
is crucial for the study of the CMB.  Before recombination, photons interacted
on short time scales with electrons via Compton scattering, so the combined
electron-proton-photon plasma was tightly coupled, moving together as a single
fluid. After recombination, photons ceased interacting with anything and traveled
freely through the universe. Therefore, when we observe CMB photons today, we
are observing the state of the cosmic fluid when the temperature of the universe
was $1/3$ eV.

Since the perturbations to the temperature field are
very small, of order $10^{-5}$, solving for the spectrum of
anisotropies is a linear problem. This means that different
modes of the Fourier transformed temperature field do not
couple with each other: each mode evolves independently.
Roughly, the large scale modes evolve very little because
causal physics cannot affect modes with wavelengths larger
than the horizon\footnote{Recall that the
horizon is the distance over which things are causally
connected.}.  
When we observe anisotropies on large angular scales,
we are observing the long wavelength modes as they appeared
at the time of recombination. Since these modes evolved little
if at all before recombination, our observations at large 
angular scales are actually of the primordial perturbations,
presumably set up during inflation\cite{inflation}. 

Inflation also set up perturbations on smaller scales, but these have
been processed by the microphysics. 
The fluid before recombination was subject to two forces: gravity
and pressure. These two competing forces set up oscillations in
the temperature\cite{husu}. A small scale mode,
begins its oscillations (in time) as soon as its
wavelength becomes comparable to the horizon. Not surprisingly, each
wavelength oscillates with a different period and phase.
The wavelength which will exhibit the largest anisotropies
is the one whose amplitude is largest at the time
of recombination. 

\begin{figure}[p]
\begin{center}
\epsfig{file=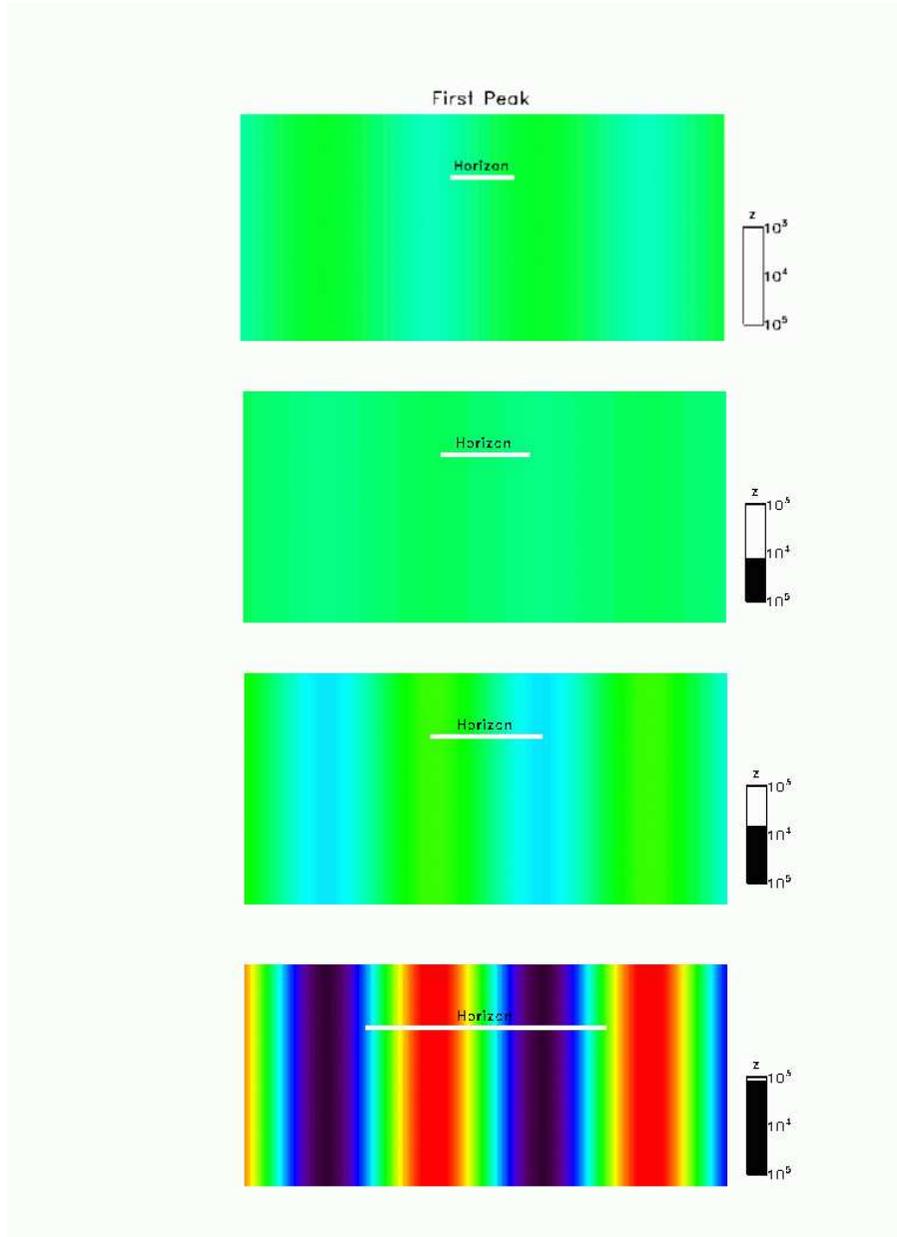, width = 12cm}
\caption{Four snapshots in the evolution of a Fourier mode.
The top panel shows the anisotropy field to this one mode
very early on, when its wavelength is still much larger
than the horizon (shown as white bar throughout). 
The second panel, at redshift $z > 10^4$ shows a time
a which the amplitude of the oscillations is very small.
THe third panel shows the amplitude getting larger; note that
the hot and cold spots in the third panel are out of phase
with those in the top panel. Finally, the bottom panel shows that
at recombination, the amplitude has reached its peak. Side
bar shows redshift ranging from $10^5$ at top to $10^3$ in
bottom panel.
}
\label{fig:peak}
\end{center}
\end{figure}

Figure \ref{fig:peak} illustrates four snapshots
in the evolution of a particularly important mode, one whose
amplitude peaks at recombination. Early on (top panel)
at redshifts 
larger than $10^5$, the wavelength of this mode was larger than
the horizon size. Therefore, little evolution took place:
the perturbations look exactly as they did when they were
first set down during inflation. At $z \sim 10^4$, evolution begins,
and the amplitudes of both the hot and cold spots decrease,
so that, as shown in the second panel, there is a time at which
the perturbations vanish (for this mode). A bit later (third
panel) they show up again; this time, the previous hot
spots are now cold spots and vise versa (compare the first
and third panels). The amplitude continues to grow until it peaks
at recombination (bottom panel).

Figure \ref{fig:peak} shows but one mode in the universe. A mode with a slightly
smaller wavelength will ``peak too soon:'' its amplitude will reach a maximum
before recombination and will be much smaller at the crucial recombination time.
Therefore, relative to the maximal mode shown in figure~\ref{fig:peak},
anisotropies on smaller scales will be suppressed. Moving to even smaller
scales, we will find a series of peaks and troughs corresponding to
modes whose amplitudes are either large or small at recombination.

\begin{figure}[thb]
\begin{center}
\epsfig{file=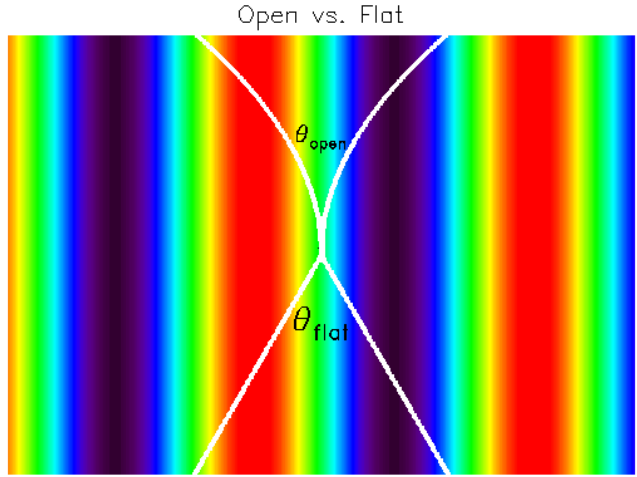, width =12cm}
\caption{Photon trajectories in an open and flat universe. The same physical
scale -- in this case the one associated with maximal anisotropy -- projects
onto smaller angular scales in an open universe because geodesics in an
open universe diverge.}
\label{fig:open}
\end{center}
\end{figure}

An important question to be resolved is at what angular scale will
these inhomogeneities show up? Consider figure~\ref{fig:open} which again
depicts the temperature field at decoupling
from the mode corresponding to the first peak. 
All photons a given distance from us will reach us today. This distance defines
a surface of last scattering (which is just a circle in the two dimensions
depicted here, but a sphere in the real universe).  
This immediately sets the angular scale $\theta$ corresponding
to the wavelength shown, $\theta \simeq$ (wavelength/distance to last scattering surface).
If the universe is flat,
then photons travel in straight lines as depicted by the bottom paths in
figure~\ref{fig:open}. In an open universe, photon trajectories diverge 
as illustrated by the top paths.
Therefore, the distance to the last scattering surface is much larger than in a
flat universe. The angular scale corresponding to this first peak is therefore
smaller in an open universe than in a flat one.

\begin{figure}[thb]
\begin{center}
\epsfig{file=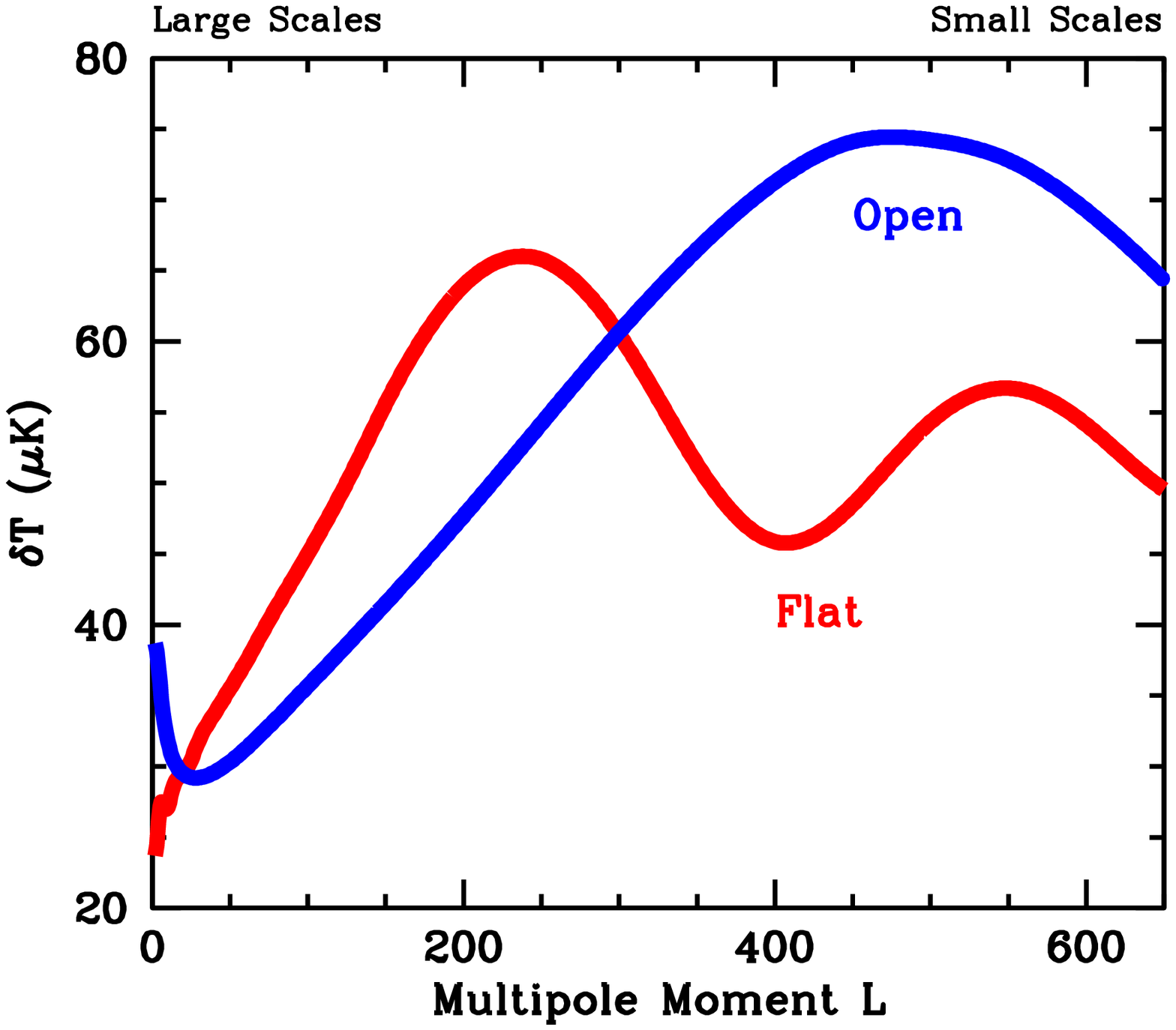, width =10cm}
\caption{The spectrum of anisotropies in an open and flat universe. Plotted is
the expected RMS anisotropy in micro Kelvin as a function of multipole moment.
The series of peaks and troughs -- the first several of which are apparent
in the flat case -- continues to small scales not shown in the plot. These are
shifted to the right in the open case, so only the first peak shows up here.
These curves are for a particular choice of cosmological parameters, corresponding
to standard Cold Dark Matter.}
\label{fig:cl}
\end{center}
\end{figure}

The spectrum of anisotropies will therefore have a series of peaks and
troughs, with the first peak showing up at larger angular scales in a
flat universe than in an open universe. Figure~\ref{fig:cl} shows the anisotropy
spectrum expected in a universe in which perturbations are set down during inflation.
The RMS anisotropy is plotted as a function of multipole moment, which is a more
convenient representation than angle $\theta$. For example, the quadrupole
moment corresponds to $L=2$, the octopole to $L=3$, and in general low $L$
corresponds to large scales. The COBE\cite{cobe} satellite therefore probed the largest
scales, roughly from $L=2$ to $L=30$. The first peak shows up at
$L\simeq 200$ in a flat universe, and we do indeed see a trough at smaller
scales and then a later peak at $L \simeq 550$. This sequence continues to
arbitrarily small scales (although past $L\simeq 1000$ the amplitudes are modulated
by damping). We also observe the feature of geodesics depicted in figure~\ref{fig:open}:
the first peak in an open universe is shifted to much smaller scales.

An important aspect of figure~\ref{fig:cl} is the accuracy of the predictions.
Although I have given a qualitative description of the evolution of anisotropies,
I and many other cosmologists spent years developing quantitative codes to
compute the anisotropies accurately\cite{codes}. This activity anticipated the accuracy with which
CMB anisotropies will be measured and therefore we strove for (i) accuracy and
(ii) speed. The former was obtained through a series of informal discussions
and workshops, until half a dozen independent codes converged to answers accurate
to within a percent. Speed is important because ultimately we will want to 
churn out zillions of predictions to compare with observations in an effort
to extract best fit parameters. Fortunately, Seljak and Zaldarriaga\cite{cmbfast} developed
CMBFAST, a code which runs in about a minute on a workstation. None of these
developments are particularly surprising: perturbations to the CMB are small, and therefore 
the problem is to solve a set of coupled linear evolution equations. The fact that
there are many coupled equations makes the problem challenging, but the fact that
these are linear more than compensates.

\section{Anisotropies: The Future}

\begin{figure}[hp]
\begin{center}
\epsfig{file=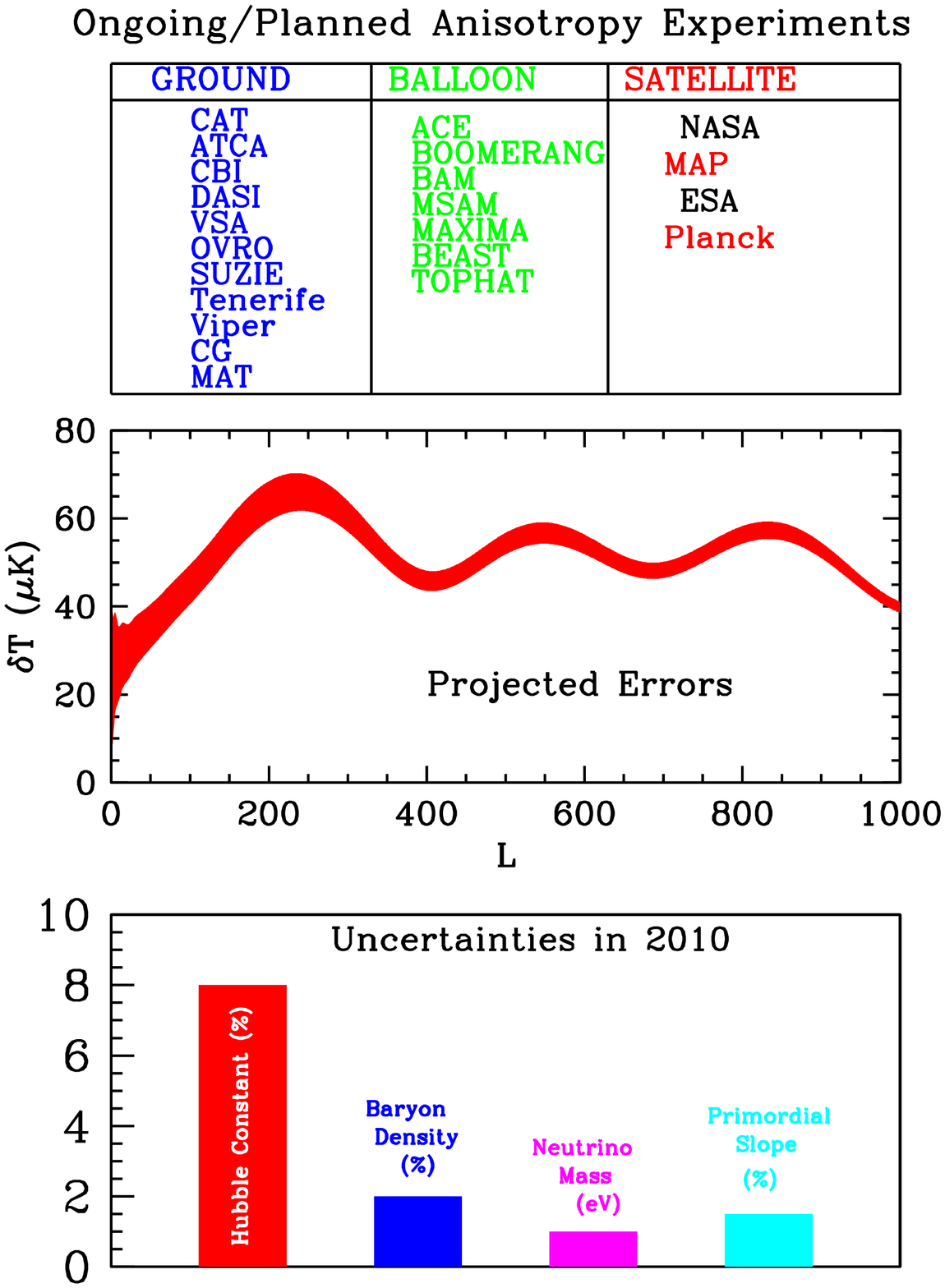, width =12cm}
\caption{The future of CMB anisotropies. {\it Top panel:} Experiments expected
to report anisotropy results within the next decade. {\it Middle panel:} Expected
uncertainty on the anisotropy after these results come in. {\it Bottom panel:}
Anticipated uncertainties in several cosmological parameters as a result of
all this information.}
\label{fig:future}
\end{center}
\end{figure}

Figure~\ref{fig:future} shows why cosmologists are so excited about
the future possibilities of the CMB. First, the top panel shows that 
people are voting with their feet. There are literally hundreds of
experimentalists who have chosen to devote their energies to measuring 
anisotropies in the CMB. Over the coming decade, this will lead to 
observations by over a dozen experiments, culminating in the efforts
of the two satellites, MAP and Planck. Some of these results are beginning
to trickle in. In particular, Viper\cite{viper}, MAT\cite{TOCO}, 
MSAM\cite{MSAM}, Boomerang NA\cite{BOOM}, and
Python\cite{PYTHON} have all reported results within the last year.

The middle panel in figure~\ref{fig:future} shows the expected errors after all
this information has been gathered and analyzed. Take one multipole moment,
at $L=600$ say. We see that the expected error is of order $5\mu$K, while the
expected signal is about $50\mu$K. At $L=600$, therefore, we expect a signal
to noise of roughly ten to one. Notice though that this estimate holds for all
the multipoles shown in the figure. In fact, it holds for many not shown in
the figure as well: it is quite possible that Planck will go out to $L\simeq 2000$.
So, we will have thousands of data points, each of which will have signal to
noise of order ten to one, to compare with a theory in which it is possible to make
linear predictions! No wonder everyone is so excited.

The final panel in figure~\ref{fig:future} shows the ramifications of getting this
much information about a theory in which it is easy to make predictions. The
exact spectrum of anisotropies depends on about ten cosmological parameters:
the baryon density, curvature, vacuum density, Hubble constant, neutrino mass,
epoch of reionization, and several parameters which specify the primordial
spectrum emerging from inflation. Figure~\ref{fig:future} shows the expected errors
in four of these parameters\cite{jungman}. In each case, all (roughly ten) other parameters
have been marginalized over. That is, the uncertainty in the Hubble constant
stated allows for all possible values of the other parameters. 

The uncertainty in the Hubble constant, of five to ten percent, comes down significantly
if one assumes the universe is flat. In any event, this uncertainty is still smaller
than the current estimates from distance ladder measurements\cite{Freedman}. 
The very small
uncertainty on the baryon density is smaller than the five percent number 
obtained by looking at deuterium lines in QSO absorption systems\cite{Burles}. 
More importantly,
the systematics involved in the two sets of determinations are completely different.
If the two determinations agree, we can be very confident that systematics are
under control. The upper limit on the neutrino mass is particularly
interesting given recent evidence for non-zero neutrino masses. The CMB alone
will not go down to $0.07$ eV, the most likely number from atmospheric neutrino
experiments\cite{atmnu}, but it will certainly probe the 
LSND region ($m_\nu \simeq 2-3$ eV)\cite{LSND}.
Further, it is possible that, in conjunction with large scale structure\cite{nulss}
 and
weak lensing measurements\cite{weaklens}, we will get to the range probed by atmospheric
neutrinos.

The final bar in the bottom panel shows the predicted uncertainty in the
slope of the primordial spectrum. While one might reasonably ask, ``What
difference does it matter if we know the baryon density or the Hubble constant
to five percent or two percent accuracy?'' the slope of the primordial spectrum
and other inflationary parameters are different. For every inflationary model
makes predictions about the primordial perturbation spectrum. The more
accurately we determine the parameters governing the spectrum, the more
models we can rule out. So it is extremely important to get the primordial
slope and other inflationary parameters as accurately as possible. These may well
be our only probe of physics at energies on the order of the GUT scale.

Along these lines, I should mention several recent developments in the field of
parameter determination. The first is an argument made by several groups for
measuring polarization\cite{polarization}. They show that accurate measurement of polarization will
decrease the uncertainty in the primordial slope by quite a bit. Even though currently
planned experiments 
may well do a nice job measuring polarization, there will still be work to
do even after Planck. So we can look forward to proposals for a next generation
experiment which measures polarization, and I believe we should strongly support such
efforts.

Another development in the field of parameter determination is
the realization that a large part of the uncertainty in some
parameters (especially some of the inflationary ones) is contributed by
treating the reionization epoch as a free parameter. In fact, it is
a function\cite{reionization} of the cosmological parameters and some 
astrophysical parameters. Recently, Venkatesan\cite{aparna} has argued that we can
use our very rough knowledge of the astrophysical parameters together
with the reionization models to reduce the errors
on the cosmological parameters.

\section{Anisotropies: The Present}

\begin{figure}[thb]
\begin{center}
\epsfig{file=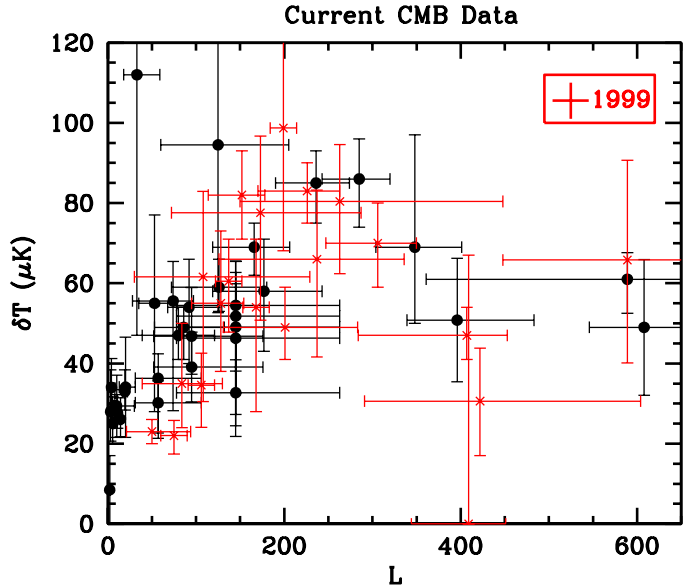, width =10cm}
\caption{Current measures of CMB anisotropy. Red crosses refer to measurements
reported within the past year. Included are all data as of November, 1999.}
\label{fig:bp}
\end{center}
\end{figure}

It is time to confront the data. Figure~\ref{fig:bp} shows all data as of
November, 1999. There are two features of this compilation worthy of
note. First, note that data reported within the last year are distinguished from
earlier results, illustrating in a very graphic way the progress of the field.
Second, figure~\ref{fig:bp} {\it understates} this progress because it was
produced before the late November release of the the Boomerang North America
``test'' flight\cite{BOOM}. Indeed, the results which follow do not include this test
flight. The papers describing the Boomerang release are fascinating if only
because one can compare the results of all data pre-Boomerang with the test
flight data. Both subsets of the data have enough power to constrain the 
curvature by themselves. They produce remarkably consistent results.

The data in figure~\ref{fig:bp} show a clear peak at around the position
expected in flat models. Indeed, a number of groups\cite{previouswork} have
analyzed subsets of this data and found it to be consistent with a flat
universe and inconsistent with an open one. I will briefly describe
my efforts with L.~Knox\cite{dk99}. We accounted for a number of facts which make it 
difficult to do a simple ``chi-by-eye'' on the data. First, every experiment 
has associated with it a calibration uncertainty: all the points from a given
experiment can move up or down together a given amount. We account for this by
including a calibration factor for each experiment and including a Gaussian prior on
this factor with a width determined by the stated uncertainties. Second, the error
bars in the plot are slightly misleading because the errors do not have a Gaussian
distribution. In particular, the cosmic variance part of the error is proportional to
the signal itself, so the error gets much larger than one would expect at high
$\delta T$. In other words, the distribution is highly skewed, with very high
values of $\delta T$ not impossible. The true distribution is close to a log-normal
distribution\cite{radcomp}, and we have accounted for this in our analysis. Finally, as alluded
to above, there are many cosmological parameters in addition to the curvature.
We do a best fit to a total of seven cosmological parameters (in addition to
eighteen calibration factors).

\begin{figure}[thb]
\begin{center}
\epsfig{file=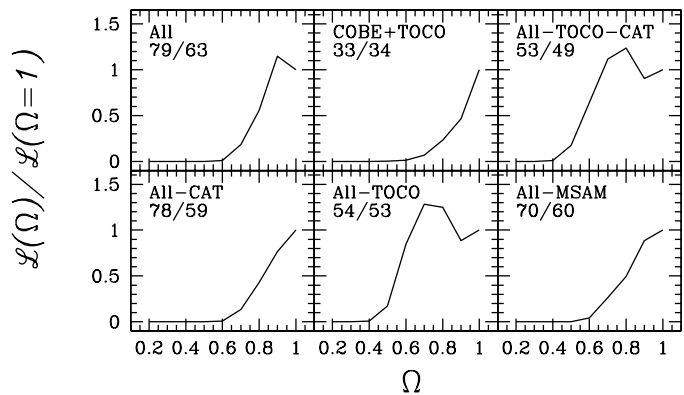, width =10cm}
\caption{Ratio of likelihood of $\Omega$ to $\Omega=1$ (flat) for different sets
of experiments. Top left panel shows results using all data; other panels
show the same ratio using only subsets of the data.}
\label{fig:chi}
\end{center}
\end{figure}

The top left panel of figure~\ref{fig:chi} shows our results. The likelihood peaks
at total density $\Omega$ very close to one (no curvature) and falls off
sharply at low $\Omega$. A universe with total density equal to $40\%$ of the
critical density is less likely than the flat model by a factor of order $10^7$.
This ratio is key because observations\cite{clusters} of the matter density in the universe
have converged to a value in the range $0.3-0.4$ of the critical density. We can combine
these two results to conclude that there must be something else besides the
matter in the universe. This conclusion probably sounds familiar to you, as 
the recent discoveries of high redshift supernovae\cite{supernovae} also strongly suggest that
there is more to the universe than just the observed matter: there is dark energy
in the universe. The exciting news is that we now have independent justification
of these results using CMB + $\Omega_{matter}$ determinations. 

\begin{figure}[thb]
\begin{center}
\epsfig{file=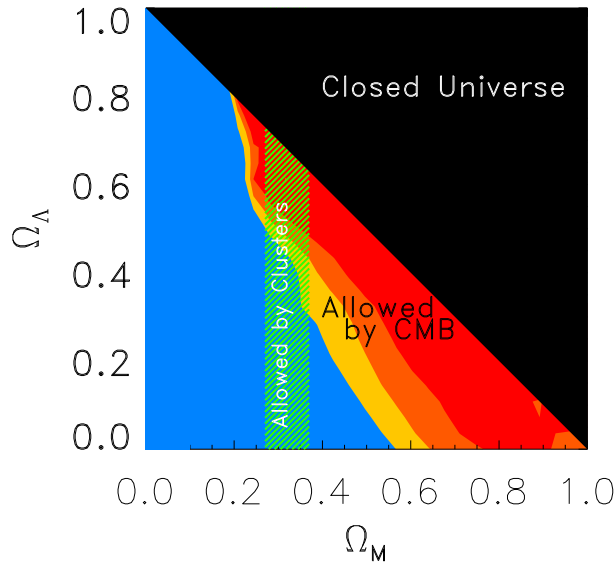, width =8cm}
\caption{Constraints on the vacuum and matter densities in the universe.
Shown are one-, two-, and three- sigma regions allowed by the CMB and 
best-fit region of the matter density from clusters.}
\label{fig:ommomv}
\end{center}
\end{figure}

One way to depict this information which has been popularized by the
supernovae teams is to plot the constraints in a space 
with vacuum energy and matter density as the two parameters. As shown in 
figure~\ref{fig:ommomv} the strongest constraints on the matter density come
from observations of baryons and dark matter in clusters of galaxies.
We obtain contours in this plane from the CMB shown in figure~\ref{fig:ommomv}.
Note that the flat line runs diagonally from top left to bottom right and is
strongly favored by the CMB. The data are so powerful that some discrimination is
appearing along this line. Very large values of $\Omega_\Lambda$ are disfavored,
and, at a much smaller statistical level, so is
$(\Omega_\Lambda=0,\Omega_{matter}=1)$. The main
result, though, is that the intersection of the regions allowed by clusters and
the CMB is at $\Omega_\Lambda \sim 0.6$, in remarkable agreement with the 
high redshift supernovae results.

This concludes my arguments for the first three claims advanced in the introduction.
Undoubtedly many of you have heard them in various forms over the past few
years. Now let's turn to the hardest claim to justify, the claim that we should
indeed believe the powerful conclusions of the CMB results. I will focus on two arguments.
First, one might be worried about the possibility that the weight of these conclusions
rests on one experiment, and one experiment might be wrong. The remaining panels of
figure~\ref{fig:chi} show that this is not a problem. We have tried removing
any one data set to see how our conclusions about $\Omega$ are affected; in all
cases, the conclusion stands. We even tried removing pairs of data
sets and again saw no change. One has to argue for a bewildering set
of coincidences if one were to disbelieve the statistical conclusions.

The second class of arguments hinges on something that was not possible until
very recently. Ultimately, skeptics will be convinced if different experiments
get the same signal when measuring the same piece of sky. Until now, this test 
has been
difficult to carry out for two reasons. First, at least at small scales, only a very small fraction
of the sky has been covered, so there has been little overlap. This has changed
a bit over the last year and obviously will change dramatically in the coming
years. Second, different experiments observe the sky differently: they smooth with 
different beam sizes and use different chopping strategies to subtract off the
atmosphere. Recently we have developed techniques which ``undo'' the experimental 
processing, thereby allowing for easy comparisons between 
different experiments\cite{MSAM}.

To illustrate the map-making technique, let us model the data $D$ in a given
experiment as
\begin{equation}
D = B T  + N
\end{equation}
where $T$ is the underlying temperature field; $B$ is the processing matrix
which includes all smoothing and chopping; and $N$ is noise which is
assumed to be Gaussian with mean zero and covariance matrix $C_N$.
To obtain the underlying temperature field $T$, we need
to invert the matrix $B$. This inversion is carried
out by constructing the estimator $\hat T$ which minimizes
the $\chi^2$:
\begin{equation}
\chi^2 \equiv (D - B \hat T) C_N^{-1} (D-B \hat T)
.\label{eq:chi}\end{equation}
We find
\begin{equation}
\hat T = \tilde C_N B C_N^{-1} D.
\label{eq:LINCOM}\end{equation}
This estimator will be distributed around the
true temperature due to noise,  where the noise covariance matrix
is
\begin{equation}
\tilde C_N \equiv <(\hat T - T)(\hat T - T) > = \left( B^T C_N^{-1} B
        \right)^{-1}.
\label{eq:NOICOV}\end{equation}

Not surprisingly, maps made from modulated data are extremely noisy.
By definition, modulations throw out information about particular modes.
For example, a modulation which takes the difference between the temperature
at two different points clearly cannot hope to say anything useful about the
sum of the temperatures. So looking at a raw, demodulated map
is a very unenlightening experience. There are two ways of getting around this noisiness
and producing a reasonable-looking map.
Before I discuss them, though, it is important to point out that even without
any cleaning up, the maps in their raw noisy states are very useful. 
They can be analyzed in the same manner as 
the modulated data, with the huge
advantage that the signal covariance
matrix is very simple to compute. Previously,
calculating the signal covariance matrix required
doing a multi-dimensional integral for every covariance element.
In the new ``map basis,'' the signal covariance matrix
simplifies to
\begin{equation}
<T_i T_j > = \sum_L {2L+1\over 4\pi} P_L(\cos(\theta_{ij}))
C_L
.\end{equation}
Indeed, one way to think of a map is that it is the linear
combination of the data for which the signal (and therefore its
covariance)
is independent of the experiment. The noise
covariance (Eq.~ \ref{eq:NOICOV}) accounts for all the experimental
processing. 

Nonetheless, we would like to produce nice looking maps, if only
to use to compare different experiments. One way to do this is
to Wiener filter the raw map, multiplying the estimator in equation
~\ref{eq:LINCOM} by $C_T (C_T + \tilde C_N)^{-1}$, which is roughly
the ratio of signal to (signal plus noise). Noisy modes are thereby
eliminated from the map\footnote{A simple way to derive this factor is
to put in a Gaussian prior in for the signal $T$, effectively adding
to the $\chi^2$ in equation~\ref{eq:chi} the term $T C_T^{-1} T$.
Minimizing this new $\chi^2$ leads to the Wiener factor.}. 

\begin{figure}[thb]
\begin{center}
\epsfig{file=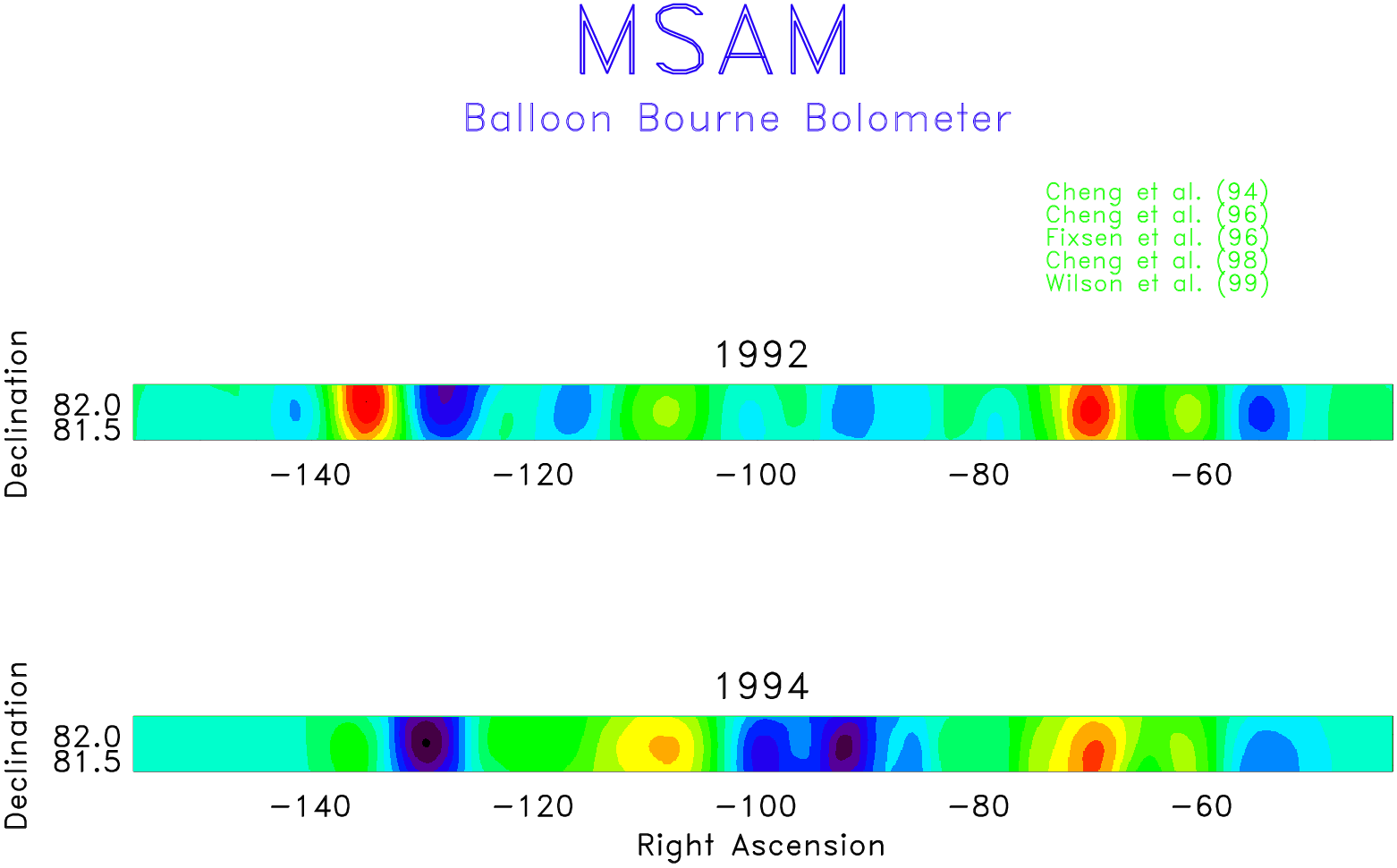,width=12cm}
\caption{Maps of two years of data from the MSAM experiment. Note that,
due to the horizontal scanning strategy, there is very little information in
the vertical direction.}
\label{fig:msam}
\end{center}
\end{figure}

An example of the Weiner filter is shown
is shown in figure~\ref{fig:msam}. The two panels are two different
years of data taken by the MSAM experiment\cite{MSAM}. It 
is well established that the two data sets are consistent~\cite{msamcomp,betatest}.
I show these because it
is important to get a sense of what constitutes good agreement.  Most
of the features are present in both experiments, but there are several --
for example the hot spot at RA $\simeq -135$ and the cold spot at 
RA $\simeq -120$ in the 1992 data -- which do not have matches. This is not surprising:
the same regions in the 1994 experiment may have been noisy so that,
in the process of throwing out the noise, the Wiener filter also
eliminated the signal. Another feature of these maps which is readily
apparent is that they only have information in one direction. There is
very little information about declination. As a corollary, the exact shapes
of the hot and cold spots in the two data sets do not agree, nor should they.
Another way of saying this is to point out that there are some modes
remaining in the maps which are noisier than others (e.g. the shapes of the
spots are noisy modes). Is there a more systematic way to eliminate noise than
the Wiener filter?

\begin{figure}[thb]
\begin{center}
\epsfig{file=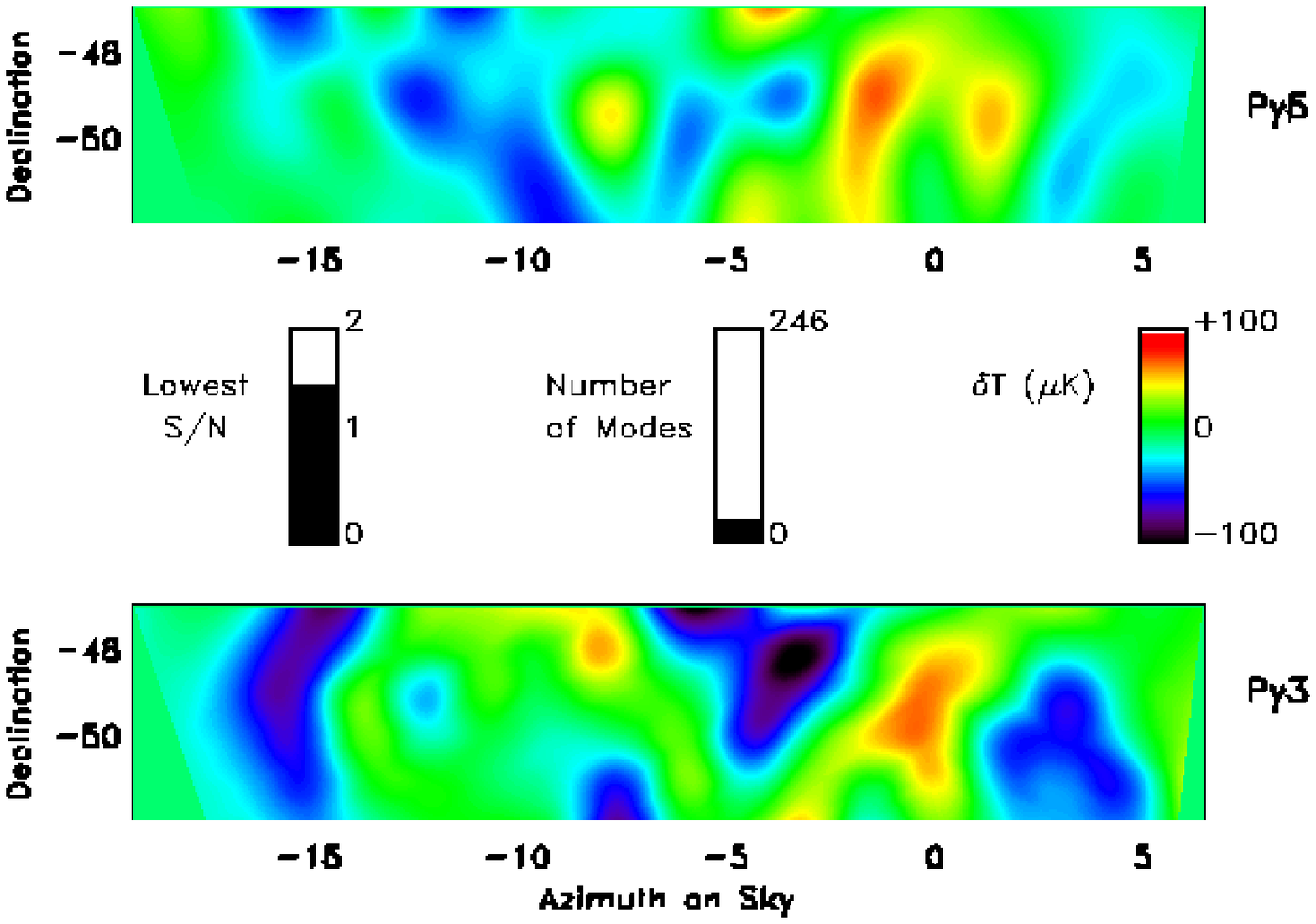,width=12cm}
\caption{Two years of Python data. Bottom panel shows data from 1995;
top panel contains much noisier 1997 data. In both cases, noisy
modes have been eliminated so that only modes with S/N
greater than $1.5$ are retained. Middle bar shows that (for
Python 97) there are of order 15 such modes out of 246 pixels
in the region.}
\label{fig:python}
\end{center}
\end{figure}

A different technique is illustrated in figure~\ref{fig:python}
in a setting which is more challenging. Whereas the two years of MSAM
data both had very high signal to noise and both were taken with
the same instrument at the same frequencies, the two years of Python
data~\cite{PYTHON,PYIII} shown were taken with completely different instruments (bolometers
in the 1995 data and HEMTs in the 1997 data) at completely different
frequencies ($90$ vs. $40$ GHz). They are therefore subject to a completely
different set of systematics and foregrounds. Further, the 1997 data
is part of a much larger region of sky covered; to get very large
sky coverage, the team sacrificed on signal to noise per pixel. 
Therefore, the signal to noise ratios of the two years are very different.

To make the maps in figure~\ref{fig:python}, I started with the raw maps and
then decomposed the data into signal to noise eigenmodes~\cite{KL}. By ordering
the data in terms of signal to noise, we can gradually and systematically
eliminate the noisiest modes. This has already been done on the 1995
in the bottom panel. The top panel contains all modes with S/N greater than
about $1.5$. As indicated by the bars, there are very few such modes,
on the order of ten. Nonetheless, many features are found in both maps.
There is the triplet of cold spots extending diagonally from $-15^\circ$
to $-10^\circ$ azimuth. There is the cold spot at $-4^\circ$ azimuth,
and the hot spot at $0^\circ$, and then finally the cold spot at the far right.
It appears to me that these two maps agree -- after far too many hours
staring at them -- as well as the MSAM maps. In fact the $\beta$ test
advocated by Bond, Jaffe, and Knox\cite{betatest} confirms this agreement.

\section{Conclusion}

The first acoustic peak in the CMB has been detected at an angular position
corresponding to that expected in a flat universe. This confirms the
fundamental prediction of inflation that the universe is flat. It also
offers independent evidence for the existence of dark energy with
negative pressure. This is but the first of many grand results
we expect to come out of the CMB over the coming decade.

\bigskip
I am grateful to my collaborators Lloyd Knox, Kim Coble, Grant Wilson,
John Kovac, Mark Dragovan, and other members of the MSAM/Python teams.
This work is supported by NASA Grant NAG 5-7092 and the DOE.

\def\Discussion{
\setlength{\parskip}{0.3cm}\setlength{\parindent}{0.0cm}
     \bigskip\bigskip      {\Large {\bf Discussion}} \bigskip}
\def\speaker#1{{\bf #1:}\ }

\Discussion

\speaker{Sherwood Parker (University of Hawaii)}
Inflation is motivated, in part, by the uniformity 
of the black body radiation coming
from places that did not have time to 
communicate since the origin of the expanding universe.
Is there any data that would exclude the following possibility:  
(1) the universe is much, and possibly
infinitely, larger than the part we can see;  
(2) the universe is much, possibly infinitely, older than 15 billion
years; and (3) there was a gravitationally driven infall of part of 
it that was reversed at a  
high energy by phenomena beyond the reach of present experiments?

\speaker{Dodelson}  It would be interesting to work out the predictions of
theories other than inflation. At present, the best alternative is topological
defects, which fare very poorly when confronted with the data. If you
can work out some prediction of your model, it would be wonderful: we
need alternatives to inflation if only to serve as strawmen. Regarding your specific
model, I don't know what you mean by larger than we can see: the standard
cosmology has this built in. If the age was much older than 15 billion years,
one would wonder why the oldest objects are roughly 10-15 billion years old.

\speaker{Jon Thaler (University of Illinois)}  
If $\Omega_\Lambda$ is 70\%\ and $\Omega_M$  is 30\%, do we still need
non-baryonic dark matter?

\speaker{Dodelson}
  Yes, due to limits from  nucleosynthesis and structure formation.

\end{document}